\documentclass[twocolumn,showpacs,preprintnumbers,amsmath,amssymb,prl]{revtex4}
\usepackage{graphicx}% Include figure files
\usepackage{dcolumn}% Align table columns on decimal point
\usepackage{bm}% bold math

\begin{document}

\preprint{Informal comment. Not for publication in a journal}% PACS, the Physics and Astronomy}

\title{A Brief Comment on the Low-Temperature Structure of LaOFeAs}

% Force line breaks with \\

\author{T. Yildirim$^{1,2}$}\email{taner@nist.gov}%
\affiliation{%
$^{1}$NIST Center for Neutron Research, National Institute of Standards and
Technology, Gaithersburg, Maryland 20899, USA
\\$^{2}$Department of Materials Science and Engineering, University of
Pennsylvania, Philadelphia, PA 19104, USA}%

\date{\today}% It is always \today, today,

\begin{abstract} 
In a recent paper [arXiv:0804.3569], Takatoshi Nomura {\it et al.} reported a structural
phase transition near 150~K in LaOFeAs and used space group "Cmma" to describe their X-ray
diffraction data. However, they did not discuss how their proposed structure compares with
the early neutron study by Cruz {\it et al.}[arXiv:0804.0795] where the low
temperature structure of LaOFeAs was described by space group "P112/n". This caused some
confusion, suggesting  that there may be some disagreement 
on the low temperature structure of LaOFeAs
as evidenced  by several inquiries that we received.  Here we show that
the proposed structures from x-ray and neutron diffraction are basically identical.
The P2/c (i.e., P112/n) cell becomes the primitive cell of the Cmma cell when the z-coordinate
of the oxygen and iron are assumed to be exactly 0 and 0.5 (these numbers were reported to
be -0.0057 and 0.5006 in neutron study).  Our first-principles total-energy calculations
suggest that the oxygen and iron atoms prefer to lie on the z=0 and 1/2 plane,
respectively, supporting Cmma symmetry. However it is more convenient to describe the
structural distortion in the primitive P2/c cell which
makes it easier to see the connection between the 
high (i.e., P4/nmm) and low temperature structures.
\end{abstract}

\pacs{Informal comment. Not for publication in a journal}% PACS, the Physics and Astronomy
\maketitle

\begin{figure}
\includegraphics[height=5cm]{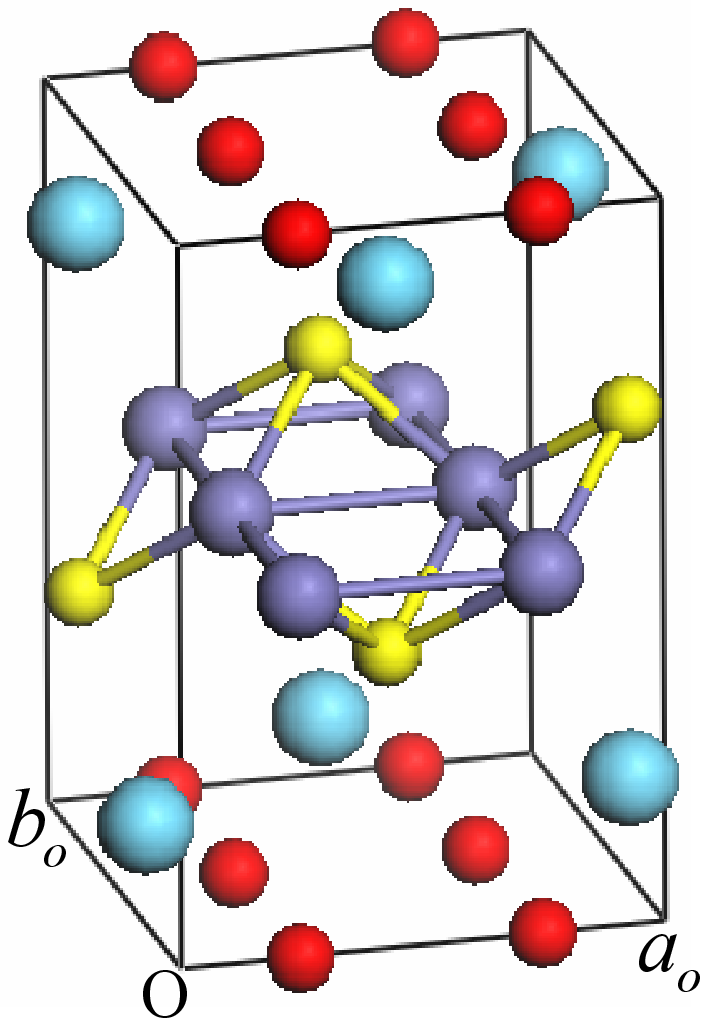} 
\includegraphics[height=5cm]{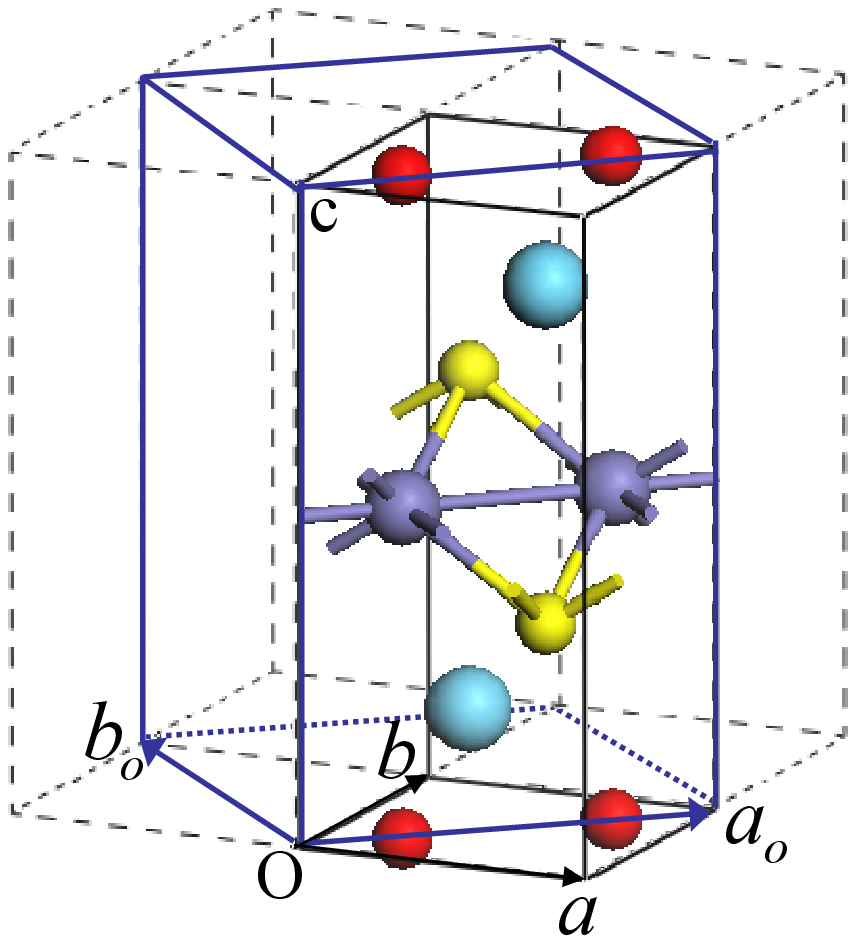} 
\caption{
(color online)
(Left) The conventional unit cell of LaOFeAs in space group Cmma 
as described in Ref.\onlinecite{hideo}.
(Right) The primitive cell of the Cmma and its relation to the conventional orthorhombic cell
(i.e. $\sqrt{2}\times\sqrt{2}$ cell). The red, light blue, purple, and yellow spheres represent
O,La, Fe, and As, respectively.
}
\label{fig1}
\end{figure}

The recent discovery of superconductivity at T$_c$'s up to 55~K in layered 
pnictide-oxide  quaternary compounds ROTmPn 
(R=La, Ce, Sm, Tm=Mn, Fe,Co, Ni, Pn=P, As)
has sparked enormous interest in this class of 
materials\cite{kamihara,sm_43k,ce_41k,pr_52k,sr_doped,dong,cruz,hideo,mazin,taner}.
In order to determine the mechanism of superconductivity in these systems,
it is very important that we understand the
structural properties of the parent compound
LaOFeAs first.

To the best of our knowledge, there have
been two experimental reports on the low temperature structure of
LaOFeAs\cite{cruz,hideo}.  The first report\cite{cruz} is from  detailed neutron powder
diffraction, which demonstrated that  LaOFeAs  undergoes an abrupt structural
distortion below $\approx 150$~K, changing the symmetry from tetragonal
(space group P4/nmm) to monoclinic space group (P112/n or P2/c) at low temperatures. 
In this neutron study, it was also reported that the system develops  
long range spin-density wave
(SDW) type antiferromagnetic ordering with a small moment and simple stripe like magnetic
structure that was first theoretically predicted\cite{dong,mazin} to occur due to Fermi surface
nesting. Our recent computational work\cite{taner} based on all-electron density functional
theory successfully explains the observed distortion quantitatively as well as the small
magnetic moment per Fe. We show that the structural distortion is closely linked to
the stripe magnetic ordering which was driven by large antiferromagnetic 
exchange interaction along
the square diagonal of the Fe-lattice and breaks the tetragonal symmetry leading to the
observed distortion\cite{taner}.

The second experimental report  on the low-temperature structure of LaOFeAs comes 
from  Prof. Hideo Hosono's group\cite{hideo}  (arXiv:0804.3569). 
The authors reported a detailed study based on 
x-ray powder diffraction. They confirmed the structural phase 
transition at 150 K in LaOFeAs as observed in the neutron 
diffraction study of Cruz {\it et al.}. However the authors 
described the low-temperature structure
by   a different space group (Cmma) from the one reported in Cruz {\it et al.}.
Unfortunately they did not compare their proposed Cmma structure  with the
P2/c structure from neutron data. This caused large confusion in the community as 
evidenced by many inquires that we received from different groups. On  initial
inspection of the x-ray study, one may think that Nomura {\it et al.} 
reached a totally different low temperature structure for LaOFeAs
than the one reported in the neutron study\cite{cruz}. However, after a close inspection
of both reports, one sees that the structures from x-ray and neutron scattering studies 
are basically identical to each other. In this brief communication, we will compare both
structures in detail to show that they are virtually identical.

\begin{figure}
\includegraphics[height=4.5cm]{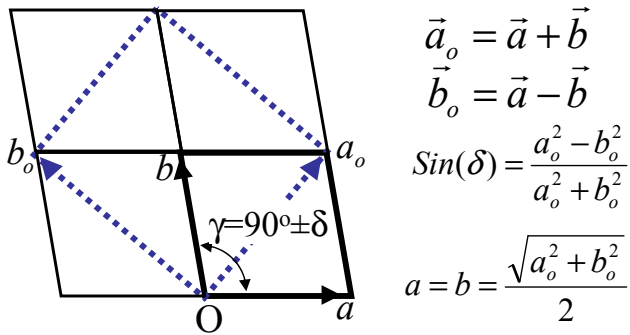} 
\caption{
(color online)
The relation between the primitive  and the orthorhombic conventional cell in Cmma. The
 equations relating the orthorhombic lattice parameters, $a_o$ and $b_o$, to the 
 tetragonal-monoclinic distortion (i.e., $\delta$) is given on the right.
}
\label{fig2}
\end{figure}

The Cmma structure proposed in Ref.\onlinecite{hideo} is shown in Fig.~1a. We note that
the orthorhombic Cmma structure is the $\sqrt{2}\times\sqrt{2}$ structure
discussed in earlier theoretical studies\cite{dong,mazin,taner}. 
This is the minimal cell to describe the
observed stripe-magnetic  ordering. Therefore, when the magnetic ordering is ignored,
it is not primitive and actually contains four LaOFeAs unit formule. 
Fig.~1b shows the primitive cell and its
relation to the conventional Cmma cell. From Fig.~1b it is clear that the primitive cell
is basically what is proposed in the neutron study\cite{cruz}. In order to make this point 
more clear, in  Fig~2, we show the relation between the cell parameters of 
the conventional orthorhombic
and primitive monoclinic cells. In particular we emphasize that the tetragonal-orthorhombic
distortion described by  Nomura {\it et al.} is basically the same
distortion as described in the neutron study. In the neutron study, this distortion was
characterized by the small deviation of the $\gamma$ angle from $90^{\rm o}$. 
Fig.~2 shows how the distortion angle is related to
the orthorhombic lattice parameters $a_o$ and $b_o$. We note that there are two identical
ways of doing the distortion; $90^{\rm o} +\delta$ or $90^{\rm o} - \delta$. Of course, 
both structures are identical and related to each other by $\pi/2$-rotation around the z-axis.
After the distortion, the Fe-Fe distance along one crystal axis gets smaller while along the other
axis it gets larger. According to our calculations\cite{taner}, the parallel aligned spins (i.e.,
the stripe direction) is the direction where Fe-Fe distance becomes shorter.
 . 

\begin{table}
\caption{The structure parameters of LaOFeAs in its low-temperature phase as obtained from
x-ray\cite{hideo} and neutron\cite{cruz} studies, respectively. 
The last column gives the differences (in absolute value) 
between x-ray and neutron structure parameters, which are very small. 
Note that the x-ray data was taken
at 120~K while the neutron data at 4~K. 
This could be one of the main reasons for the tiny difference
between these two structures. The difference in z-values listed in the last column are obtained by using the 
difference in fractional coordinates from the neutron and the x-ray data times the c-value from the neutron data.}
\begin{center}
\begin{tabular}{|c||c|c||c|c|} \hline \hline
 &\multicolumn{2}{|c||}{X-ray  Cmma (120 K) } & Neutron (4K) & \\ 
 & Conventional &  Primitive  &  P112/n   & $|$ Diff. $|$ \\ \hline\hline
$a$ & 5.68262 \AA & 4.02806 \AA & 4.0275 \AA& 0.0006 \AA\\
$b$ & 5.71043 \AA & 4.02806 \AA & 4.0275 \AA& 0.0006 \AA\\
$c$ & 8.71964 \AA & 8.71964 \AA & 8.7262 \AA& 0.0066 \AA\\
$\gamma$  &   $90^{\rm o}$ & $89.7203^{\rm o}$ & $ 90.279^{\rm o}$ & \\
$|\delta |$ &  0 &  $ 0.2797^{\rm o}$ & $0.279^{\rm o}$  & $0.0007^{\rm o}$\\
La(z) & 0.14171 & 0.14171 &  0.1426  & 0.0078 \AA\\
As(z) & 0.65129 & 0.65129 & 0.6499 & 0.0121 \AA\\
O(z)  & 0  & 0 & -0.0057 & 0.0497 \AA \\
Fe(z) & 0.5 & 0.5  &  0.5006  & 0.0052 \AA \\ \hline\hline
\end{tabular}
\end{center}
\end{table}

After having presented the relation between the conventional and primitive cell of the
Cmma, we translate the x-ray structure from Cmma into its primitive form in Table~1
 and compare it with the
neutron structure with P2/c symmetry. 
From this table, it is clear that both structures are basically identical.
The main difference between the x-ray and neutron-data is the z-values of the oxygen and
iron ions, which dictates what the symmetry of the final structure will be. 
Since the numerical difference for these coordinates are very small, this is a 
technical point rather than
quantitative difference in the structures. When the z-values of oxygen and iron are
exactly 0 and 0.5, then the symmetry is higher with space group Cmma and the neutron structure
becomes the primitive cell of the x-ray structure.  
These values were reported to be 
0.5006 and 0.0057 in the neutron paper\cite{cruz}, 
which are essentially no different from 0.5 and 0. 
Hence technically the Cmma and P112/n structures are identical.

A possible explanation for the tiny difference between the structures listed in Table~1 is
that the x-ray data was taken at T=120 K while the neutron
data is taken at T=4~K. The  $c$-lattice parameter from x-ray and neutron studies seem to be
slightly different also. This could be due to  oxygen defects that may be present in the samples.
At any rate, the structural parameters as well as the z-values of the atomic positions are
in excellent agreement with each other.

As a final note, we report what the structural optimization from density functional theory gives.
As discussed in detail in Ref.\onlinecite{taner}, both the magnetic and crystal structure are 
successfully obtained from our calculations. For example, we obtained a=5.66803 \AA, b=5.73383
\AA, and c=8.70417 \AA $\;$ for the conventional cell, all in excellent agreement 
with the experimental data. Similarly, the
z-values (in fractional coordinates) for the La and As are obtained to be 0.138518 and 0.65155,
which are again in excellent agreement with the data listed in Table~1. Finally, our structural
optimization for the z-values of the oxygen and Fe ions always give the high symmetry locations,
i.e. 0 and 0.5, respectively, supporting the Cmma symmetry.

In summary, there is no  difference in 
the structure described in x-ray\cite{hideo} and neutron\cite{cruz} studies.
Technically it seems that Cmma is the correct  symmetry where the oxygen and iron ions are at the
high symmetry sites (i.e., z=0 and 0.5, respectively). However it is easier to see the
connection between the high-temperature tetragonal phase and the low-temperature distorted 
structure if one works in the primitive cell of Cmma.

The author acknowledges useful discussions with 
J. Simmons, Q. Huang and J. W. Lynn.

\end{document}